\documentclass[12pt,preprint]{aastex}

\newcommand{\ca}{\ion{Ca}{2}}
\newcommand{\mg}{\ion{Mg}{2}}
\newcommand{\oi}{\ion{O}{1}}
\newcommand{\ci}{\ion{C}{1}}
\newcommand{\kms}{km s$^{-1}$}
\newcommand{\mum}{$\mu$m}

\begin{document} 

\title{Low Carbon Abundance in Type Ia Supernovae}

\author{G. H. Marion, P. H\"oflich, J.C. Wheeler and E. L. Robinson}
\affil{Astronomy Department, University of Texas at Austin, 
       Austin, TX 78712, USA}

\author{C. L. Gerardy}
\affil{Astrophysics Group, Imperial College London
       Blackett Laboratory, Prince Consort Road, London, SW7 2AZ, U.K.\\
       and The University of Texas at Austin}

\and

\author{W. D. Vacca}
\affil{SOFIA-USRA,
       NASA Ames Research Center,   
       MS N211-3,   
       Moffett Field, CA 94035-1000}

\begin{abstract}
We investigate the quantity and composition of unburned material in the outer layers of three normal Type Ia supernovae (SNe Ia): 2000dn, 2002cr and 2004bw. Pristine matter from a white dwarf progenitor is expected to be a mixture of oxygen and carbon in approximately equal abundance.  Using near-infrared (NIR, $0.7-2.5$ \mum) spectra, we find that oxygen is abundant while carbon is severely depleted with low upper limits in the outer third of the ejected mass. Strong features from the \oi\ line at $\lambda_{rest} = 0.7773$ \mum\ are observed through a wide range of expansion velocities $\approx 9 - 18 \times 10^3$ \kms.  This large velocity domain corresponds to a physical region of the supernova with a large radial depth.  We show that the ionization of C and O will be substantially the same in this region.  \ci\ lines in the NIR are expected to be $7-50$ times stronger than those from \oi\ but there is only marginal evidence of \ci\ in the spectra and none of \ion{C}{2}.  We deduce that for these three normal SNe Ia, oxygen is more abundant than carbon by factors of $10^2 - 10^3$.  \mg\ is also detected in a velocity range similar to that of \oi.  The presence of O and Mg combined with the absence of C indicates that for these SNe Ia, nuclear burning has reached all but the extreme outer layers; any unburned material must have expansion velocities greater than $18 \times 10^3$ \kms.  This result favors deflagration to detonation transition (DD) models over pure deflagration models for SNe Ia.

\end{abstract}


\section{Introduction}
There is general agreement that Type Ia supernovae (SNe Ia) result from the combustion of a degenerate carbon/oxygen (C/O) white dwarf (WD) star as first predicted by \citet{hoyle60}.  The currently accepted scenario for SNe Ia is the explosion of a C/O WD near the Chandrasekhar mass ($M_{Ch}\approx 1.44M_{\odot}$). The explosion is triggered by compressional heating near the center of the WD.  

A C/O WD is the final evolutionary stage for stars with a main sequence mass of $\approx 3-8 M_{\odot}$. Initially the progenitor has insufficient mass to produce a SN Ia.  To reach $M_{Ch}$, the WD must accrete additional matter through Roche-lobe overflow from a main sequence, red giant, or WD companion.   Quasi-steady shell burning of accreted material in conditions of low pressure and high temperature produces a mixture of C and O with a mass ratio near unity \citep{dominguez01}.  If two WDs are entirely mixed in a merger scenario \citep{webbink84}, the C to O ratio by mass will also be close to 1:1 \citep{benz89}.

One of the keys to understanding the explosion is the propagation of the nuclear flame.  Two modes of burning can be distinguished: {\em detonation}, with burning velocities slightly above the speed of sound, and {\em deflagration}, with burning front velocities well below sound speed.  The two models for SNe Ia currently under discussion are Pure Deflagration \citep{ivanova74, nomoto76} and Delayed Detonation Transition (DD) \citep{khokhlov91}.  Despite some differences of the numerical treatment of nuclear burning fronts by different groups, both models share the same initial sequence: nuclear burning begins near the center of the WD and the burning front becomes Rayleigh-Taylor unstable.  These instabilities result in large scale mixing of burned and unburned matter.  During the early burning phase, energy is deposited into the WD causing the outer layers to expand with velocities close to the speed of sound \citep{hk96}.  Classical deflagration models cease burning in this phase and leave a chemical structure characterized by a thick layer of unburned C and O surrounding inner regions of intermediate-mass and iron-peak elements.

Recent refinements with multi-dimensional deflagration models predict that, within the context of standard models, a substantial mass of unburned material is mixed toward the center of the SN and burning products are mixed outward toward the surface.  Primordial and processed elements intermingle throughout the radial chemical structure rather than forming distinct concentric layers \citep{Gamezo03, Ropke05}.  Significant quantities of material remain unburned because the deflagration burning front is unable to catch up to the expansion of the WD near the surface \citep{hk96, khokhlov01}.  Thus the outer layers in current deflagration models remain unburned by means of causality and large pockets of unburned C/O will be found at various depths because of the nature of the Rayleigh-Taylor instabilities.  These generic properties of pure deflagration models seem to be in contradiction to observations of so called ``Branch-normal'' SNe Ia \citep{Branch93} that require vanishing amounts of unburned matter and radially layered chemical structures \citep{Barbon90, pah95, Wheeler98, fisher98, Hamuy02, m03}. 

DD models also begin with subsonic deflagration burning, but they make a transition to a supersonic detonation front.  For models of normal SNe Ia, the detonation eradicates most of the chemical inhomogeneities, producing a radially stratified chemical structure with very little unburned matter left at the surface.  We note that questions remain about how the deflagration to detonation transition occurs.  For recent reviews of SN Ia physical models see \citet{Branch99}, \citet{Hill00}, and \citet{pah05}.

In this paper, we investigate the presence of unburned material in the outer layers of three normal SNe Ia by examining carbon and oxygen features in near-infrared (NIR, 0.7 - 2.4 \mum) spectra.  The NIR is an excellent region for this search due to the presence of several strong lines from C and O ions.  We use observational evidence to make estimates for the relative abundance of O and C in these SNe Ia.  General methods are used for this analysis to produce upper limits on the abundance of these elements and avoid dependence on specific models or parameters.  Section 2 describes the observations including data acquisition and reduction as well as details of the supernovae.  Section 3 presents results with line identifications and measurements. Section 4 provides methods and calculations for C and O locations, ionizations, line strengths, and abundance ratios.  Section 5 provides a summary and our conclusions.

\section{Observations}
\subsection {Data Acquisition and Reduction}

Low and medium resolution NIR spectra were obtained from SNe Ia using the 3.0 meter telescope at the NASA Infrared Telescope Facility (IRTF) with the SpeX medium-resolution spectrograph \citep{Rayner03}.   SpeX provides single exposure coverage of the wavelength region $0.8-2.5$ \mum.  Using a grating and prism cross-dispersers (SXD mode), the spectral resolution of the instrument is $R=750-2000$ and with a single prism (LRS mode) the resolution is $R=150-250$.  A serendipitous characteristic of the LRS observing mode is that it extends spectral coverage at the blue end of the NIR to about 0.65 \mum.  The additional wavelength region includes \oi\ at $\lambda_{rest} = 0.7773$ \mum\ enabling comparison of absorption features from this line to other NIR spectral features.

Individual exposure times are capped at 150s to limit background noise from OH.  A typical set is limited to ten exposures for a total of 25 minutes integration time in order to remain within timescales of atmospheric variability.  The source is nodded along the slit using an A\_B\_B\_A\_A\_B\_B\_A\_A\_B pattern.  Calibration images are obtained from A0V standard stars.  Each observation set also includes calibration images from internal lamps for flat fielding and wavelength calibrations.

The data were reduced using a package of IDL routines specifically designed for the reduction of SpeX data (Spextool v. 3.2; \citeauthor{Cushing04} 2004). These routines perform pair subtraction, flat-fielding, aperture definition, spectral tracing and extraction, residual sky subtraction, host galaxy subtraction, and wavelength calibration for data acquired in both the prism mode and the cross-dispersed mode.  Corrections for telluric absorption were performed using the extracted spectrum of an A0V star and a specially designed IDL package developed by \citet{Vacca03}. These routines generate a telluric correction spectrum by comparing the spectrum of an A0V star to a model A0V spectrum that has been scaled to the observed magnitude, smoothed to the observed resolution and shifted to the observed radial velocity. The telluric correction spectrum is then shifted to align the telluric absorption features seen in the SN spectrum and divided into the target spectrum.  

The individual spectra from each object are combined using using median statistics.  Noise levels are calculated and recorded by Spextool for each wavelength bin.  Because the data are subject to bad pixels which can skew the combined mean value and the error on the mean, we substitute the median and median absolute deviation (MAD) as robust estimates of these statistics.  MAD is defined as: $1.4826 \times med(|(f_{\lambda}-med)|)$ where $f=$ flux, $med$ is the median of the combined values, and 1.4826 is a constant introduced on the assumption of a Gaussian distribution of initial values.  The {\it error on the median} at each wavelength is equal to the one-sigma noise level and given by $\frac{MAD}{\sqrt{N}}$, where $N$ is the number of spectra that have been combined.

Detailed procedures to obtain and reduce IRTF data from SNe Ia are described by \citet{m03}. 

\subsection{The Supernovae}
SNe 2000dn, 2002cr, and 2004bw were identified as Type Ia from optical spectra, using the \ion{Si}{2} feature at $\lambda_{rest} =6355$ \mum\ which defines the class.  The observed characteristics of spectral evolution, light curve, and maximum luminosity indicate that each of these supernovae are, within uncertainties, Branch-normal SN Ia events.

Spectra from these objects were obtained using the LRS mode of the SpeX instrument at $R=250$. Each of the spectra contain a prominent feature from the \oi\ line at $\lambda_{rest} = 0.7773$ \mum. The spectrum from SN 2000dn is found in \citet{m03} while the spectra from SNe Ia 2002cr and 2004bw have not previously been published.

{\bf SN 2000dn} was observed on October 2, 2000. A relative B-band light curve for SN 2000dn provided by W. D. Li and M. Papenkova of the Lick Observatory Supernova Search team indicates that $B_{max} =$ October 6. We estimate the date of $V_{max}$ to be $B_{max} +2$ days, making $V_{max} =$ October 8.  The recession velocity of the host galaxy (IC 1468) reported by LEDA is 9599 \kms\ which  makes it a very distant object to observe with SpeX.  The estimated $M_V = -19.0$ mag is slightly dimmer than the canonical value for SNe Ia of -19.3 mag, but the decline rate of 0.8  \citep{m03} suggests that the luminosity estimate may be low.  This appears to be a normal SN Ia near the lower end of the decline rate distribution.

{\bf SN 2002cr} was observed on May 8 \& 10, 2002, under adverse weather conditions.  The spectrum contains data only from May 10 because the May 8 data is not useful below 0.8 \mum.  \citet{Matheson_7891} estimate that $V_{max} =$ May 8 from optical spectra.  A second estimate by \citet{Meikle_7891} using optical spectra finds the date of $V_{max}$ to be May 9 or 10.  The Sternburg Supernova Catalog lists $V_{max}$ as May 13 without reference.  The development of \mg\ features in this spectrum is consistent with other NIR spectra obtained a few days before $V_{max}$ from normal SNe Ia with well-documented light curves.  NIR spectral development favors the May 10 date for $V_{max}$.  The recession velocity of the host galaxy (NGC 5468) reported by LEDA is 2875 \kms.  Amateur photometry indicates $M_V \approx -19.1$ mag which is consistent with a normal SN Ia.  We note that this spectrum appears to be shifted in wavelength space from other spectra in our sample.  We have reviewed our wavelength calibrations and conclude that this shift is real and is an example of diversity in normal SNe Ia.

{\bf SN 2004bw} was observed on May 31, 2004.  The LRS observing mode was chosen for this object to maximize signal-to-noise.  \citet{Foley_8353} estimate the date of $V_{max}$ to be June 6.  The NIR spectral development is characteristic of SNE Ia about a week before maximum which agrees with the estimate by \citeauthor{Foley_8353}. The recession velocity of the host galaxy (MCG +00-38-19) is reported by LEDA to be 6418 \kms.  Photometry is sparse for this object and amateur observations provide a wide range of possible values for $M_V$ from -18.8 to -19.5 mag.    Despite the sparse data, all indicators (optical and NIR spectra plus $M_V$ estimates) agree that this object is a normal SN Ia with a detached high velocity component in the NIR \ca\ triplet (see discussion \S3.6).

\section{Results}

\subsection{The Spectra}
\begin{figure}
\epsscale{0.8}
\plotone{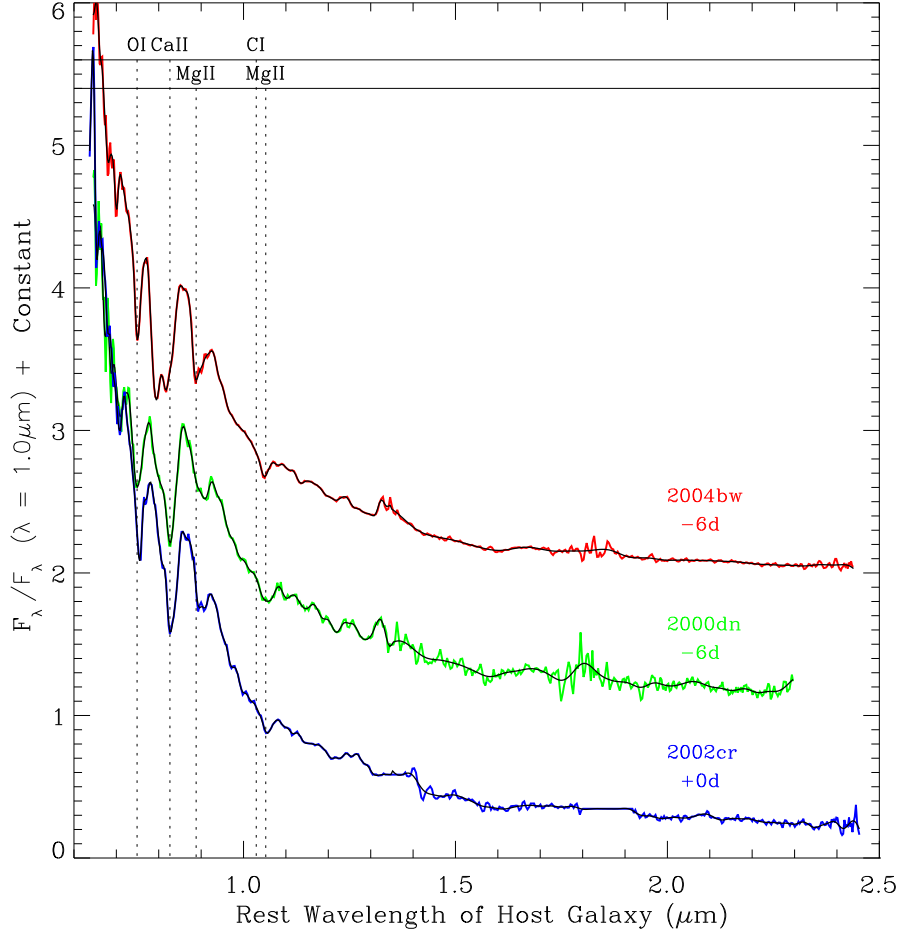}
\caption{Three NIR spectra from normal SNe Ia acquired at the IRTF using the SpeX instrument in LRS (single prism) mode, $R=250$.  The spectra have been normalized to unity at 1.0 \mum\ and shifted by a constant.  Superimposed on the raw spectra are the same spectra after smoothing with a Fourier Transform.  The locations of prominent absorption features are identified by a vertical dotted lines at expansion velocities of 12,000 \kms\ for the \mg\ lines and 11,000 \kms\ for \ci, \oi, and \ca.  No features are evident from the \ci\ line even though it is expected to be $\approx 50$ times stronger than the \oi\ line.\label{3full}}
\end{figure}

LRS spectra from SNe Ia 2000dn, 2002cr, and 2004bw are presented in Figure \ref{3full}.  They are normalized to unity at 1.0 \mum\ and shifted by a constant for clarity.  The Doppler shifted positions for \oi\ at 0.7773 \mum, the \ca\ ``infrared triplet'' with mean value of 0.8579 \mum\ and \ci\ at 1.0693 \mum\ are marked in the figure with vertical dotted lines at 11,000 \kms.  The shifts for \mg\ at 0.9227 and 1.0927 \mum\ are marked at 12,000 \kms.  \oi, \ca, and \mg\ produce strong absorption features in these spectra, but there is no evidence for \ci.  

Superimposed on the raw spectra in Figure \ref{3full} are the same spectra after smoothing with a Fourier Transform.  A detailed description of the procedure and parameters for smoothing are found in \citet{m06}.  The smoothed spectra for these three SNe Ia are used for all subsequent figures in this paper.

Measured Doppler shifts are used to estimate expansion velocities.  In regions where the lines are formed close to the photosphere, as they are near the absorption minima, projection effects make the Doppler shift estimates lower limits on expansion velocities (\S 4.1). 

We note that the spectrum from SN 2002cr is shifted slightly in wavelength relative to the other spectra.  This spectrum has Doppler velocities that are 1,000 - 1,500 \kms\ lower than other SNe Ia in our sample.  In all other respects, this spectrum is similar to the others indicating that the chemical structure is approximately the same.  This spectrum was obtained at an epoch 5-6 days after the two spectra so the Doppler velocities are expected to be slightly lower.  Small velocity differences are indications of diversity within the class of normal SNe Ia.

Inspection of Figure \ref{3full} shows that the energy flux from SNe Ia drops off rapidly with increasing wavelength.  Within a steeply sloping region, the deepest part of an absorption feature may not correspond to the wavelength of the maximum departure from the continuum.  This is not a significant effect where the continuum is relatively flat, as in the region of the \ion{Si}{2} line at 6355 \mum, but it can potentially affect features in the NIR by $\approx 0.005$ \mum\ or $\approx$ 1,500 \kms.  To remove this influence, we fit a local continuum in the region of each absorption feature and normalize the spectrum to a flat continuum before identifying the wavelength of absorption minimum.

\begin{figure}
\epsscale{0.8}
\plotone{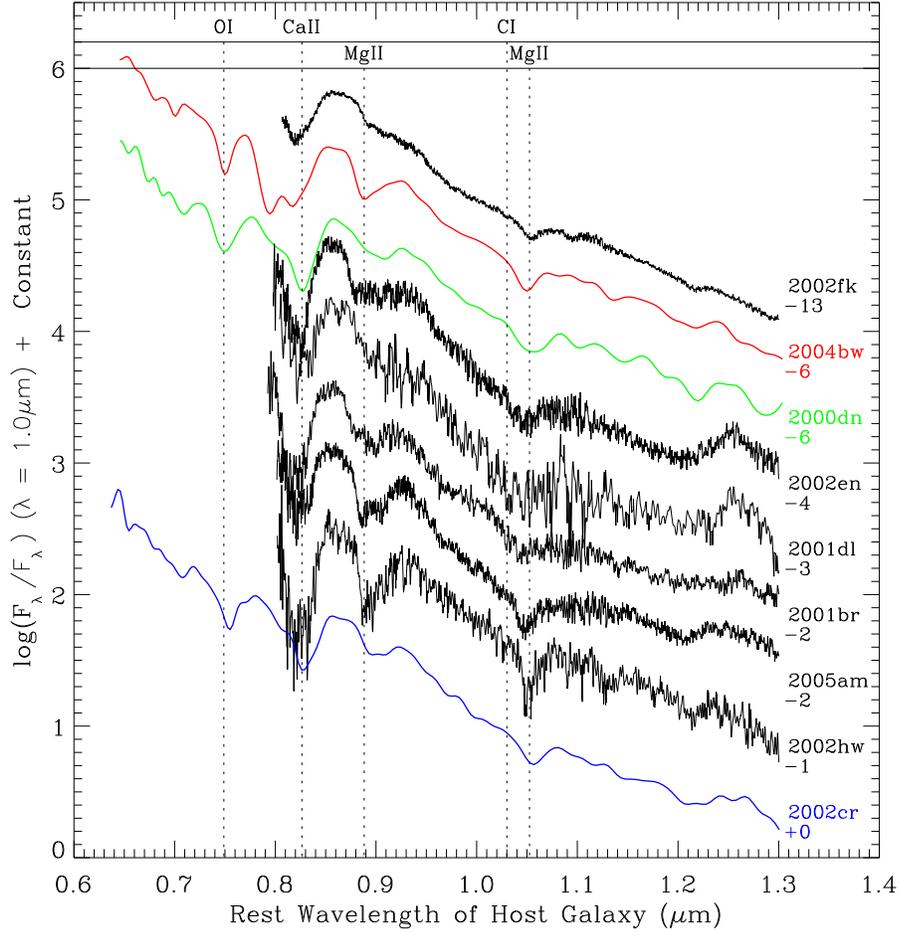}
\caption{Plot of the log of the flux for nine spectra from normal SNe Ia obtained at or before $V_{max}$. The spectra have been normalized to unity at 1.0 \mum\ and the vertical displacement is arbitrary.  The LRS spectra are conspicuous compared to the SXD spectra due to their high S/N and the fact that they extend beyond 0.8 \mum.  The SXD spectra have not been smoothed.  The locations of prominent absorption features are identified by a vertical dotted lines.  \oi\ is detected in each of the LRS spectra near 0.75 \mum.  All spectra in this sample contain the \ca\ feature observed near 0.82 \mum, and the two \mg\ features observed near 0.89 and 1.05 \mum.  No features are evident from the \ci\ line in any of the spectra  even though it is expected to be $\approx 50$ times stronger than the \oi\ line.\label{9j}}
\end{figure}

Figure \ref{9j} shows the same three LRS spectra along with six SXD spectra from normal SNe Ia to demonstrate the similarities in spectra from normal SNe Ia at similar epochs.  The SXD spectra in this figure have not been smoothed.  The spectra are normalized as in Figure \ref{3full} and the wavelength region is reduced to enlarge the region under discussion.   The LRS spectra are conspicuous due to their high S/N and the fact that they extend to wavelengths shorter than 0.8 \mum.  All nine spectra exhibit strong absorption features from \ca\ observed near 0.82 \mum\ and \mg\ observed near 0.89 and 1.05 \mum. The LRS spectra show \oi\ near 0.75 \mum.  There is no obvious detection of \ci\ in the spectra. 

Six of the nine spectra in Figure \ref{9j} were published earlier by \citet{m03}.  The exceptions are the spectrum from SN 2002cr and both spectra from SN 2005am (obtained at -5 and -0 days before $V_{max}$) which have not been previously published.

\subsection{\oi\ Features}
The locations of the four strongest \oi\ lines in the NIR are displayed in Figure \ref{3j} at a Doppler velocity of 11,000 \kms.  To help with line comparisons, a condensed version of Tabel \ref{colines} (\S 4.3) gives rest wavelengths, shifted wavelengths and line strengths normalized to the strongest line from each ion.  A strong absorption feature from the 0.7773 \mum\ line is found near 0.75 \mum\ in each of the LRS spectra.  This is the only \oi\ line detected in these spectra, but the other lines are expected to be significantly weaker with estimates of 30\%, 6\%, and 2\% of the 0.7773 \mum\ line strength. In addition, the lines at 0.8446 and 0.9264 \mum\ are close to much stronger features from \ca\ and \mg\ respectively.  The non-detection of \oi\ lines other than 0.7773 \mum\ is easily explained by weak lines and obscuration by other features.  

\begin{figure}
\epsscale{0.7}
\plotone{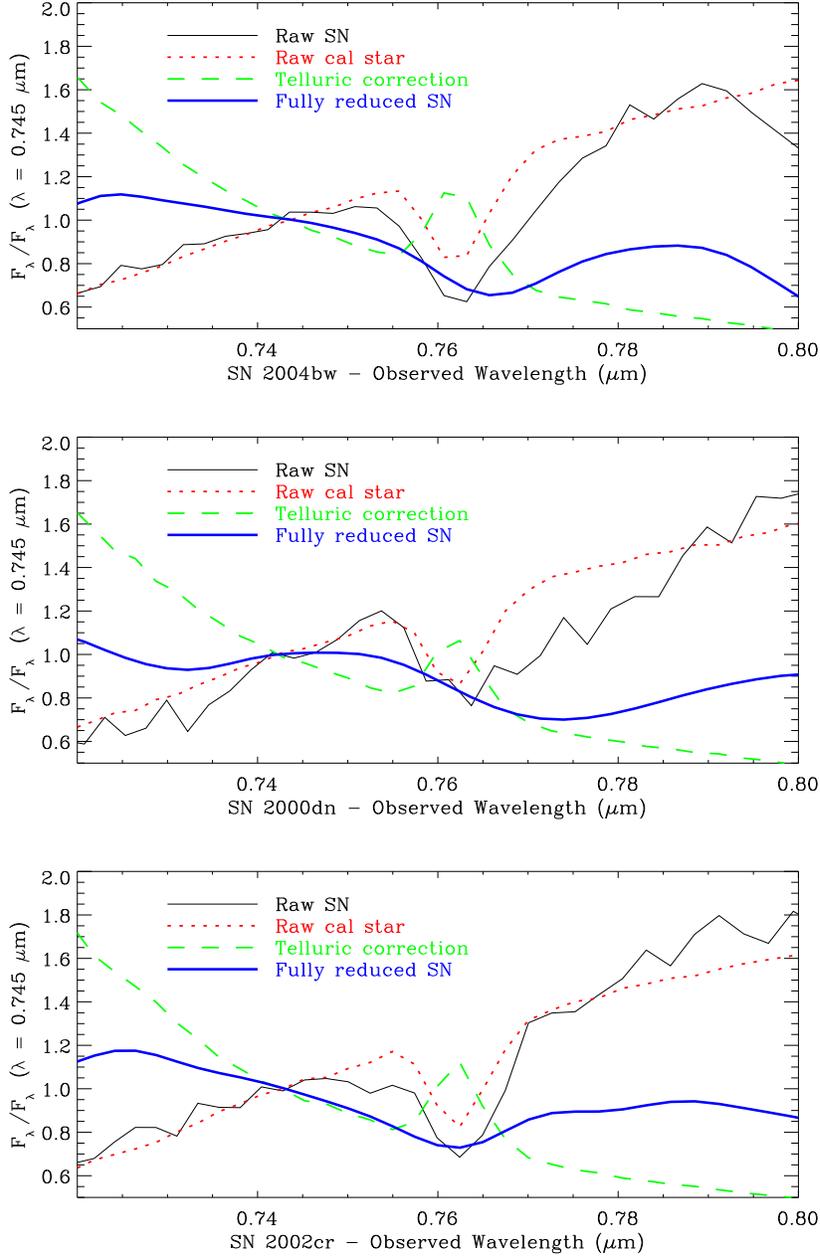}
\caption{Removal of telluric absorption in the feature produced by \oi\ at 0.7773 \mum. All spectra are normalized to 1.0 at 0.745 \mum.  Wavelength in this figure is as observed and not shifted to the rest wavelength of the host galaxy as in the other figures.  The absorption feature in the calibration star (dotted line) is clearly smaller than the absorption from the SN (solid line).  The telluric correction spectrum is the dashed line and the final reduced and smoothed spectrum is the thick solid line. The absorption feature in the final spectrum appears to be weaker in this figure than in other figure because the wavelength region is small compared to the flux region.\label{otell}} 
\end{figure}

Atmospheric absorption from \oi\ at 0.7773 \mum\ will contaminate the signal from the SNe.  Figure \ref{otell} shows that the size of the telluric feature is much smaller than the observed absorption in the supernova spectrum.  Wavelength in this figure is as observed and not shifted to the rest wavelength of the host galaxy as in the other figures. The spectra in Figure \ref{otell} are normalized to 1.0 at 0.745 \mum.  The absorption feature in the calibration star (dotted line) is clearly smaller than the absorption from the SN (solid line).  The telluric correction spectrum is the dashed line and the final reduced and smoothed spectrum is the thick solid line.  The absorption feature in the final spectrum appears to be weaker in this figure than it appears in other figures because the wavelength region is small compared to the flux region.

\begin{figure}
\plotone{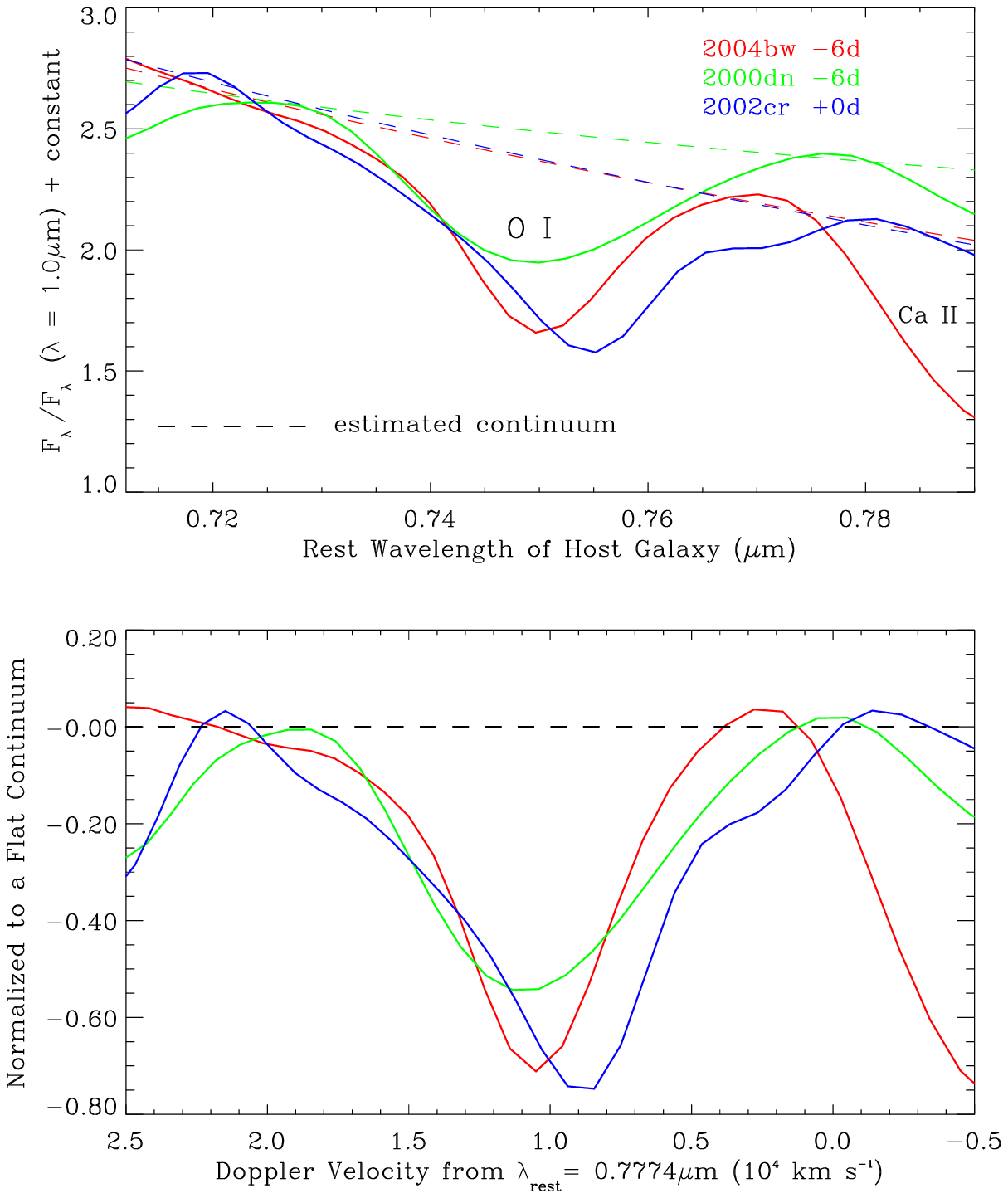}
\caption{Enlarged view of the region around the absorption feature from \oi\ at 0.7773 \mum.  The top panel shows the \oi\ features in three LRS spectra with estimated locations for the continua indicated by dashed lines that are colored to match the spectra. In the lower panel, the same line profiles are normalized to a flat continuum and are plotted in velocity space determined by the Doppler shift from the rest position of the line.  The line forming region for \oi\ lies between the absorption minimum and the farthest extent of the blue wing.  The location of the line forming region for \oi\ is similar to that of \mg\ (Figure \ref{mg}).\label{oi}}
\end{figure}

The absorption features from \oi\ at 0.7773 \mum\ are expanded in Figure \ref{oi}.  The top panel shows line profiles from the LRS spectra with estimated locations for the continua indicated by dashed lines which are colored to match the spectra.  The sizes and shapes of these features are generally similar with the most notable difference being the truncated bottom in the spectrum of SN 2000dn.  The shallowness of this feature is likely due to noise.  One-sigma noise in the region of this feature is $0.07-0.10$ for SN 2000dn, $0.04-0.08$ for SN 2002cr, and $0.04-0.06$ for SN 2004bw.

In the lower panel, the same line profiles are normalized to a flat continuum and plotted in velocity space rather than wavelength space.  Expansion velocities are determined by the Doppler shift from the rest position of the \oi\ line.  The Doppler velocities of the absorption minima range from $9 - 11 \times 10^3$ \kms\ with the blue wings crossing the detection limit at $18-20 \times 10^3$ \kms.  Thus the depth of the line forming region in velocity space is $\approx 9 \times 10^3$ \kms\ for \oi\ in these spectra.

\subsection{\ci\ Features}
\begin{figure}
\plotone{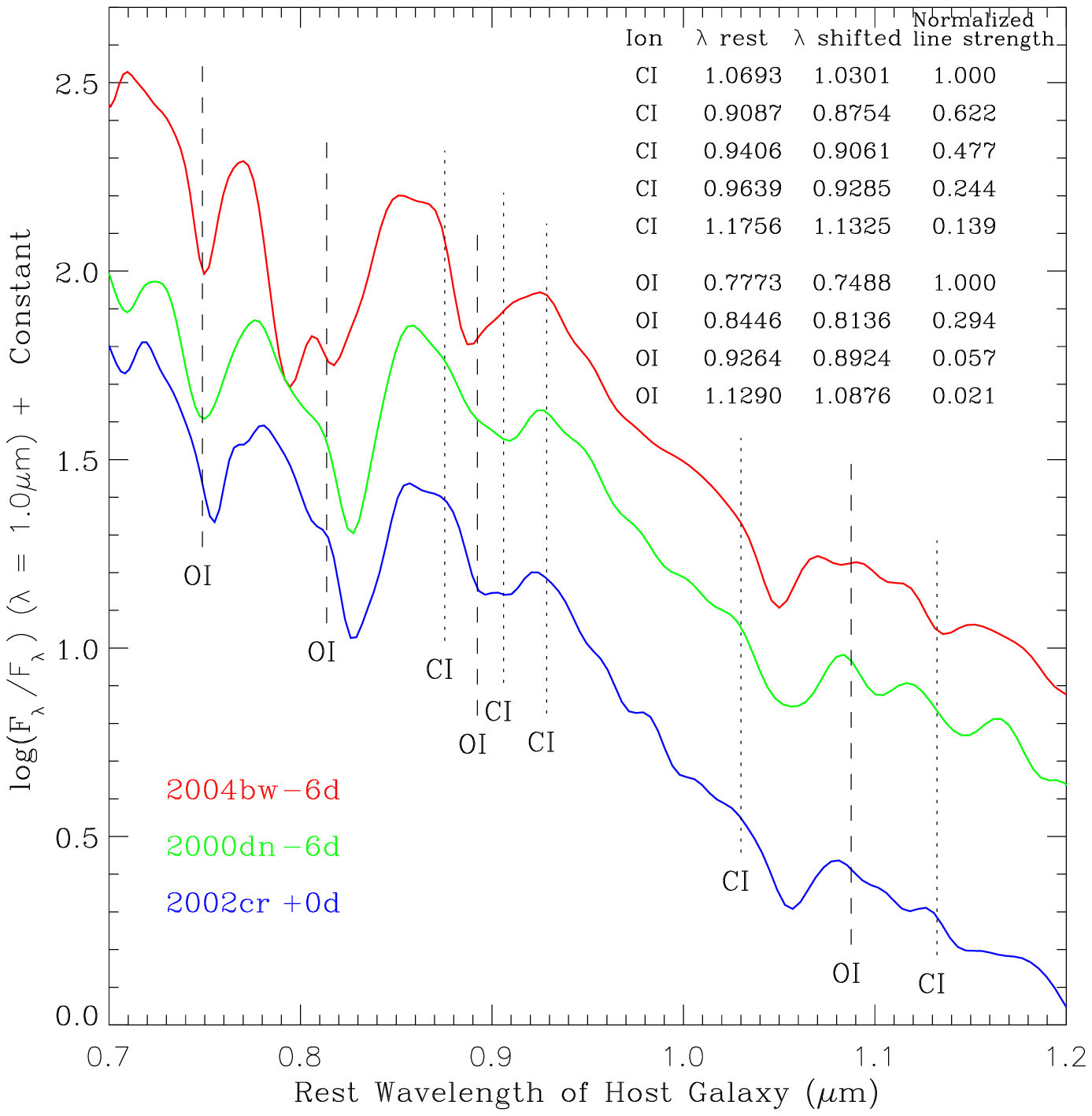}
\caption{The strongest NIR lines from \oi\ and \ci\ are shifted by 11,000 \kms\ and plotted with the LRS spectra. 9,000 \kms\ is $\approx 0.01$ \mum\ on the red side of the marked position of each line and 18,000 \kms\ is $\approx 0.03$ \mum\ on the blue side.  To help with line comparisons, a condensed version of Tabel \ref{colines} is provided in the top right corner. The rest wavelengths, Doppler shifted wavelengths at 11,000 \kms, and line strengths normalized for each ion are given for strong \oi\ and \ci\ lines.  \oi\ from $\lambda_{rest}=0.7773$ \mum\ is found in all spectra near 0.75 \mum.  No reliable detections are made in these spectra for any other lines from \oi\ or \ci.\label{3j}}
\end{figure}

The locations the five strongest \ci\ lines in this region are plotted in Figure \ref{3j} at a Doppler velocity of 11,000 \kms.  The strongest \ci\ line in the NIR at $\lambda_{rest}=1.0693$ \mum\ is expected to be almost 50 times stronger than the \oi\ line at 0.7773 \mum\ (given equal abundance and departure from LTE, Tabel \ref{colines}), but there are no strong features from this line in Figure \ref{3j} or in the nine spectra displayed in Figure \ref{9j}.  A few possible weak detections seen in Figure \ref{3j} are actually smaller than one-sigma noise.  The best detection candidate is the feature in the spectrum of SN 2004bw found near 1.135 \mum\ and possibly due to the 1.1756 \mum\ line at an expansion velocity of 10,500 \kms.

\begin{figure}
\plotone{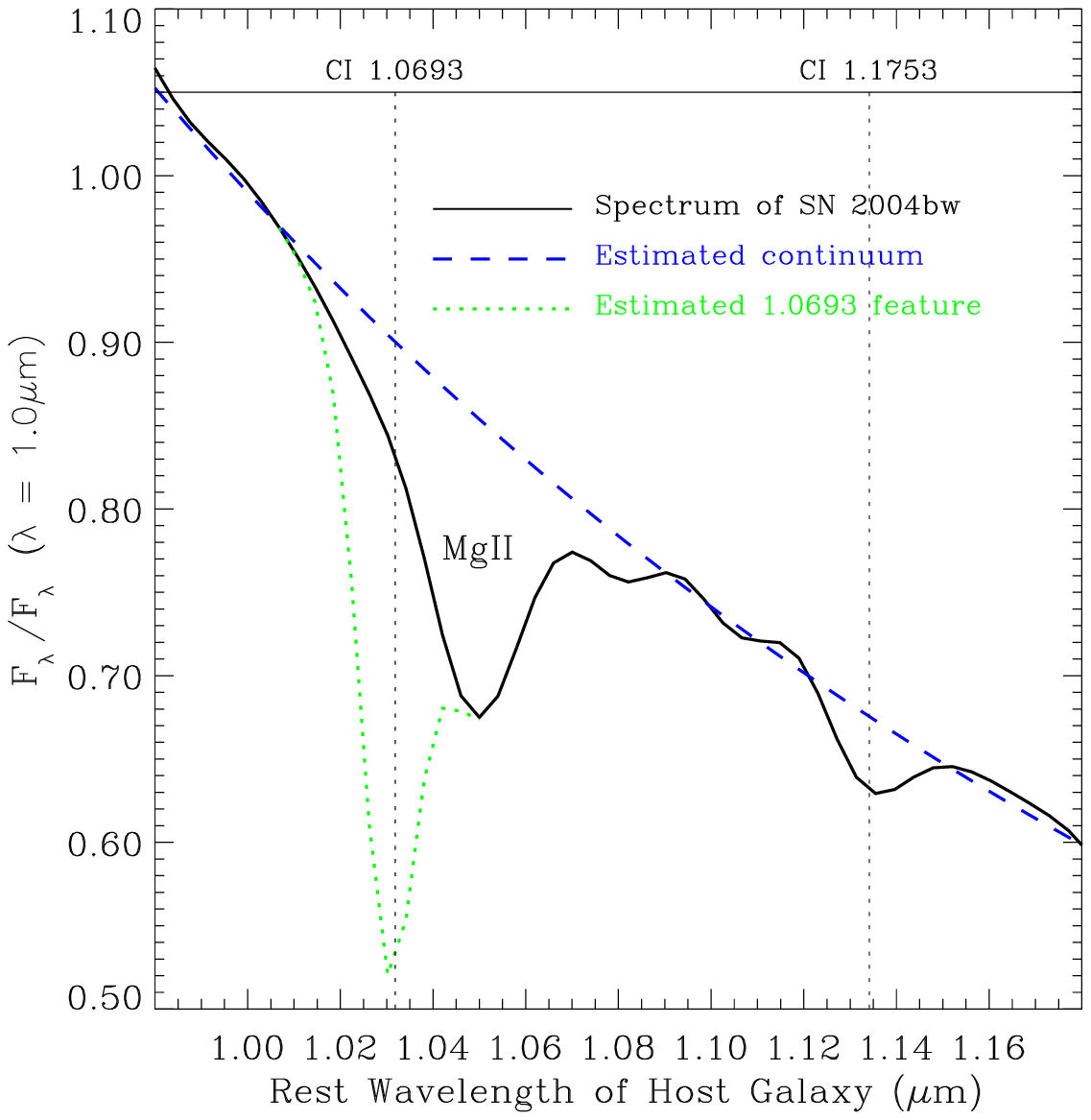}
\caption{The spectrum from SN 2004bw with the location of \ci\ lines from 1.0693 and 1.1756 \mum\  marked at a velocity of 10,500 \kms.  The large absorption feature with a minimum near 1.05 \mum\ is due to \mg.  The spectrum is plotted as a solid line and the estimated location of the continuum is plotted with a dashed line.  If the feature found near 1.135 \mum\ were due to \ci, then there should also be an enormous absorption feature near 1.03 \mum\ as indicated by the dotted line in the figure.\label{noci}}
\end{figure}

Figure \ref{noci} shows the spectrum from SN 2004bw in the wavelength range $0.98-1.18$ \mum\ with \ci\ lines from 1.0693 and 1.1756 \mum\ marked at 10,500 \kms.  We use this figure to investigate both the most likely \ci\ detection and the strongest \ci\ line in our data.  The large absorption feature with a minimum near 1.05 \mum\ is due to the \mg\ triplet with mean rest wavelength of 1.0927 \mum.  The spectrum is plotted as a solid line, the estimated continuum is plotted as a dashed line, and the one-sigma noise level is plotted with dash-dot lines.  Noise levels increase in region $1.08-1.15$ \mum\ due to a reduction in atmospheric transmission by about 25\%.  The actual region of increased opacity is from $\approx 1.1-1.16$ \mum, but in the figure the entire spectrum is shifted to the rest wave length of the host galaxy. 

We note that the estimated line strength of the 1.0693 \mum\ line is 7.2 times stronger than the line strength of the 1.1756 \mum\ line.  If the feature found near 1.135 \mum\ is due to \ci\ at 1.1756 \mum\ producing a line depth of 0.045, then the line from \ci\ at 1.0693 \mum\ should produce an absorption feature with a depth of 0.324.  A synthetic feature of this size appears in the figure as a dotted line with absorption minimum near 1.03 \mum.  The presence of such a feature is clearly ruled out and we infer that the weak feature at 1.135 \mum\ is not due to \ci.

We can extend this argument to all \ci\ lines in Figure \ref{3j}.  If any of the other possible detections of other \ci\ lines were real, then the 1.0693 \mum\ line would be expected to produce a prominent feature.  No such feature is observed in any of the spectra in our sample.  In addition, the Doppler velocities of the possible detections within individual spectra vary by $2-5 \times 10^3$ \kms.  We conclude that none of these features are due to \ci.

Neutral carbon has been detected in subluminous SN Ia 1999by \citep{pah02}.  The presence of carbon in this supernova is interpreted to indicate that burning took place at lower temperatures than for normally bright SNe Ia.  \citet{ger04b} also find \ci\ in NIR spectra from Type Ib/c supernovae.  These reports indicate that neutral carbon can be easily detected in spectra from SNe Ia.  They reinforce the conclusion that the absence of \ci\ features in our data is significant.

\subsection {\ion{C}{2} and \ion{O}{2}}

There is no evidence for \ion{C}{2} in these spectra. \citet{Branch03} and \citet{Garavini05} report the identification of \ion{C}{2} features from the line at 0.6580 \mum\ in optical spectra of normal SNe Ia.  Unfortunately that wavelength region is not accessible with the spectra in our sample.  There are few \ion{C}{2} lines in the NIR and the minimum excitation levels are much higher than for \ci\ (Tabel \ref{colines}).  Lines from different ionization stages cannot be directly compared, but these very high excitation levels suggest that \ion{C}{2} lines will require significantly greater abundances than \ci\ to produce detectable features.  Even at the upper limit of expected temperatures, \ion{C}{2} lines in the NIR remain significantly weaker than the 0.6580 \mum\ line for which the detection is reported.   

There is no evidence in the spectra for \ion{O}{2}.  Minimum excitation levels for the \ion{O}{2} lines are greater than 25 eV which is well above the $9 - 11$ eV for the \oi\ lines and make it unlikely that a detectable signal will be produced in the region covered by these spectra.  

\subsection{\mg\ Features}
The strongest NIR lines from \mg\ are 0.9229 and 1.0927 \mum.  All spectra in Figure \ref{9j} exhibit absorption features from both lines, near 0.89 and 1.05 \mum\ respectively, with Doppler velocities near 12,000 \kms.  No convincing detections are made for \mg\ from lines at 0.7890, 0.8228, or 1.0092 \mum.  

\begin{figure}
\plotone{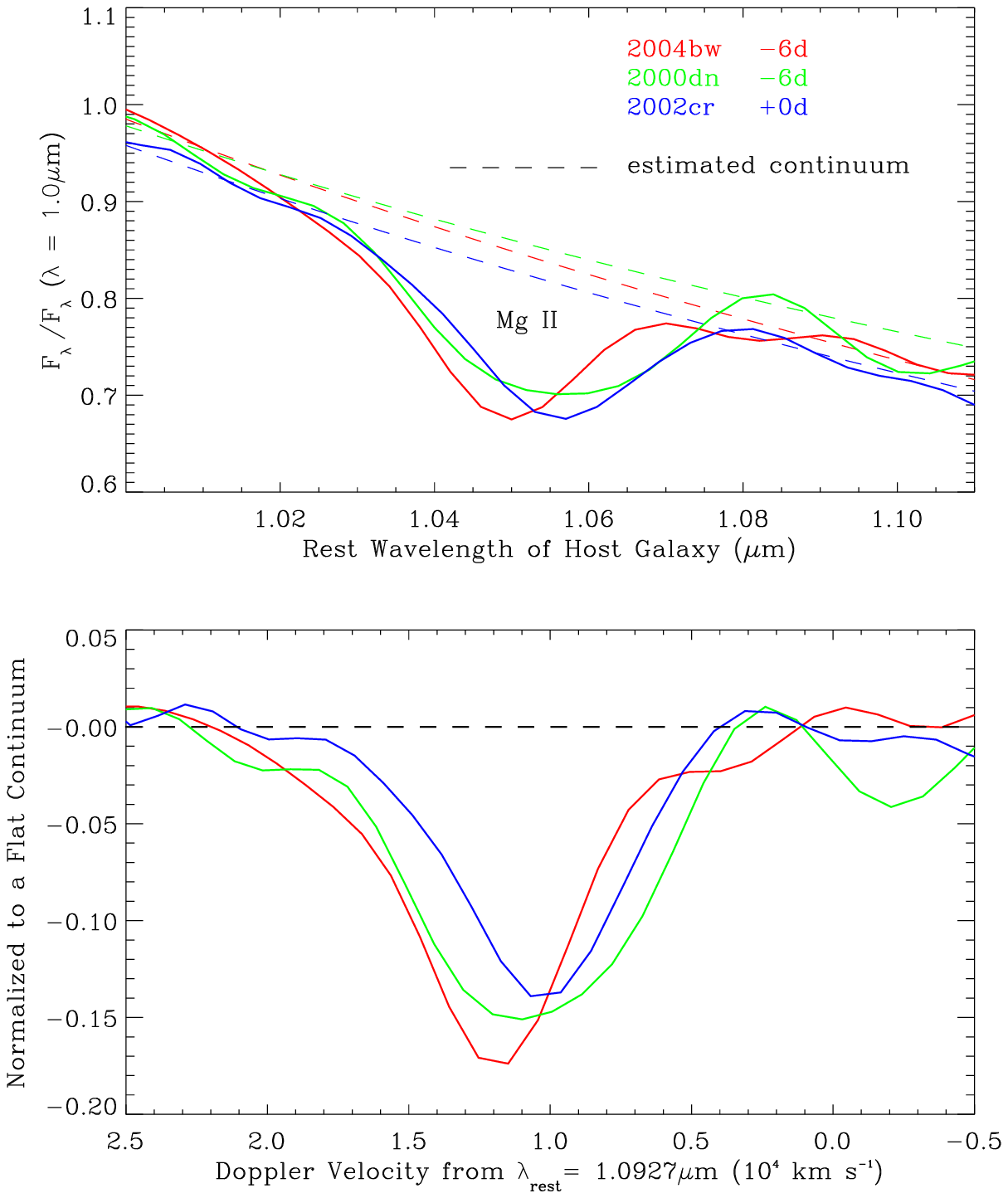}
\caption{Enlarged view of the region around the absorption feature from \mg\ at 1.0927 \mum.  The top panel shows the \mg\ features in three LRS spectra with estimated locations for the continua indicated by dashed lines that are colored to match the spectra. In the lower panel, the same line profiles are normalized to a flat continuum and are plotted in velocity space determined by the Doppler shift from the rest position of the line.  The line-forming region for \mg\ lies between the absorption minimum and the farthest extent of the blue wing.  The location of the line-forming region for \mg\ is similar to that of \oi\ (Figure \ref{oi}).\label{mg}} 
\end{figure}

The top panel of Figure \ref{mg} shows the LRS spectra in the region of the absorption feature produced by the \mg\ line at 1.0927 \mum.  The estimated location of the continuum for each spectrum is marked by a dashed line in the same color as the corresponding spectrum.  The profiles from this line are generally similar, with the spectrum from SN 2000dn displaying a double bottom that is probably due to noise.  One-sigma noise in this region is $0.02-0.03$ for SN 2000dn, $0.005-0.01$ for SN 2002cr, and $0.01-0.015$ for SN 2004bw.

The lower panel shows the same \mg\ features plotted in velocity space and normalized to a flat continuum.  The velocity minima are in the range $10-13 \times 10^3$ \kms\ and the blue wings reach $20-21 \times 10^3$ \kms.  The depth of the line forming region for \mg\ in these spectra is $\approx 10 \times 10^3$ \kms\ which is large enough to make nearly all of the magnesium singly ionized (\S 4.2).  The velocity range for \mg\ is very close to the velocity range we identified for \oi.

\subsection{\ca\ Features}

The spectral region  $0.7 - 1.2$ \mum\ is dominated by a huge P Cygni feature produced by the \ca\ infrared triplet with a weighted mean rest wavelength of 0.8579 \mum.  The very low excitation value of this line (1.7 eV, Table \ref{mgcalines}) makes it strong enough to obscure the signal from most nearby lines.  Doppler velocities of the absorption minima are between 11,000 and 13,000 \kms. 

Many SNe Ia show a detached high velocity component in the absorption feature produced by the \ca\ IR triplet \citep{Hatano99, Li01, wang03, Mazzali05a, Mazzali05b, Quimby05}.  The spectrum of SN 2004bw clearly shows a detached high velocity \ca\ feature at $\approx$ 22,000 \kms\ (Figure \ref{3j}).  \citet{ger04a} have have argued that this high velocity feature is the spectral signature of a region where the shock from the explosion encounters and compacts circumstellar material forming a dense shell.  The shell is composed of material from the circumstellar environment, possibly including an accretion disk, as well as the extreme outer layers of the supernova ejecta. The disk has a high scale height giving it a large covering factor that makes it observable from a wide range of angles \citep{Quimby05}.  \citeauthor{ger04a} provide a figure showing the mass of the shell as a function of expansion velocity.  Our measurement of $v_{exp} \approx 22,000$ \kms\ corresponds to $\approx 2 \times 10^{-2} M_{\odot}$ for the mass of the shell. 

\section{Limits on Carbon Abundance}

\subsection{The Location of Unburned Material}
If the O features observed in these spectra are due to unburned material, then C should occupy the same physical space.  Alternatively, oxygen can be produced by explosive carbon burning in regions where the temperature was sufficient to burn C but not hot enough to consume O.  In that case, magnesium is expected in the same physical region as O because Mg is also a product of carbon burning.  Since oxygen-rich domains may exist in SNe Ia from both burned and unburned sources, it is possible that C will not be found in all areas where O is detected.

Oxygen has been detected in these spectra with Doppler velocities of $\approx 9-18 \times 10^3$ \kms, measured between the absorption minimum and the blue wing at the detection limit. The full range of Doppler velocities is required to identify the entire line forming area and not to be restricted to the region closest to the photosphere. The red half of an absorption feature does not provide useful information because it is contaminated by emission.  All lines contain an emission component due to a non-vanishing line source function.  

Expansion velocity in SNe Ia is directly proportional to radial distance as a result of homologous expansion.  A continuous range of expansion velocities can therefore be used to represent contiguous physical space in the supernova.  Velocities uncertainties from spectral resolution are $\approx \pm 1,200$ \kms\ but Doppler velocities can be determined to $\approx \pm 500$ \kms\ by interpolating between pixels.

We note that the continuum forming region is extended.  If the line-formation region for a specific ion is close to this photosphere, many incoming photons are absorbed from matter with a projected velocity of $v \times sin(\theta )$ where $\theta < \pi/2$.  This projection effect shifts the observed position of absorption minima to the red which causes the true expansion velocity to be underestimated by $10-20$\%. An example is the \oi\ line-forming at the epochs considered here.  For lines formed far outside in the last scattering radius for continuum photons, the photosphere appears to be more point-like and the projection effect is minimized. Typical examples are the NIR \mg\ lines at later times when the Mg-rich layers are well above the photosphere.

An estimate of the relationship between expansion velocity and mass distribution has been calculated for DD models by \citet{pah02}.  All DD models release about the same amount of energy because nearly the entire C/O WD is burned, so the velocity/mass relationship is not sensitive to specific model parameters.  \citeauthor{pah02} determine that about half of the mass in a SN Ia is exterior to a velocity of 9,000 \kms\ but only about 5\% of the mass is beyond 18,000 \kms.  This means that the line-forming region for \oi\ in the SNe Ia we have observed extends to the extreme outer layers and includes more than one third of the total mass of the supernova.  The lack of carbon in this region implies that 5\% is the upper limit on the mass remaining unburned with C and O in approximately equal abunance.  

This estimate, that $\approx$5\% of the mass has a velocity exceeding 18,000 \kms, is an upper limit based on the assumption that nearly all of a C/O WD progenitor has been burned.   If a different model is used, for instance a pure deflagration model, then less mass is burned and less energy is released.  The reduction in total kinetic energy implies that even less mass will have expansion velocities exceeding 18,000 \kms.

\subsection{Finding the Dominant Ionization}
In SNe Ia, transitions from one ionization stage to another are rapid in both time and velocity space due to the dominance of radiative transitions over collisional processes.  Model calculations for radiative transfer in SNe Ia envelopes indicate that UV opacities in the ejecta are dominated by Fe-group lines that create a large optical depth for UV photons.  The distribution of the photo-ionization spectrum is determined by Fe opacities that generate a smooth quasi-continuum.  The ionization balance for other elements is hardly dependent on their own photo-ionization boundaries.  As a result, elements with similar ionization potentials cover similar spatial and velocity ranges in models for SNe Ia, with boundaries defined by Fe transitions \citep{pah95}.  

Table \ref{ip} shows that the first ionization potential for carbon is 11.26030 eV and for oxygen, 13.61806 eV.  These values are between the first ionization (7.9024 eV) and the second ionization (16.1878 eV) of iron.  The energy required to doubly ionize both C and O is greater than the ionization edge for \ion{Fe}{3}.  Fe opacities restrict both carbon and oxygen to be primarily neutral in the physical zone bounded by Fe ionizations at 7.9 eV and 16.1 eV. 

Small differences in the ionization potential between two elements will create a narrow band in radial space where one element is ionized and the other is not.  The size of this region is model dependent but it is on the order of $1-2 \times 10^3$ \kms\ \citep{pah95} which is much smaller than the observed depth of the \oi\ line forming region $\approx 9 \times 10^3$ \kms.  This large radial depth eliminates the possibility that within the line forming region a significant amount of carbon will exist in an ionization state different from oxygen.

Elements with a first ionization potential similar to magnesium (7.64624 eV) will also be singly ionized in the velocity range observed for \mg.  Table \ref{ip} shows atoms that fit that description include: Si, Ca, Ti, Mn, Fe, Co, and Ni.

\subsection{Estimating Line Strength in LTE}
Level populations in SNe Ia are far from local thermodynamic equilibrium (LTE), however the line strength of transitions from elements with similar ionization and excitation levels show similar departures from LTE so they can be directly compared \citep{pah95}.  Uncertainties due to differences in the departure coefficients are a factor of 2-3 which does not alter our conclusions.  The use of LTE for calculating line strengths removes the dependence of our estimates on specific SN Ia model parameters or methods.

Temperatures in the line producing region at this epoch are expected to be between 5,000 - 10,000K \citep{hmk93, kmh93}.  Line strength is calculated for each line at temperatures of 5,000K and 10,000K using $(gf)*10^{-\Theta (eV)}$ where $\Theta = 5040/T$.  The results are given in Tables \ref{colines} and \ref{mgcalines}.   To facilitate comparison of line strengths, the computed values at 5,000K are normalized to the strongest line for each ion.  Rest wavelengths are given in air.  Many of the indicated wavelengths are a mean value, weighted by oscillator strength, for two or more lines. For these multiplets, the $\log(gf)$\ value and excitation minima are listed for the strongest line in the blend.

Temperature estimates can be derived from the line strength estimates in combination with observations of \mg\ features.  The spectra in Figure \ref{9j} show strong features from \mg\ lines at 0.9229 and 1.0927 \mum\ but no detections from lines at 0.7890, 0.8228, or 1.0092 \mum.  These results are consistent with relative line strengths calculated for a temperature of 5,000K.  Calculations for the same lines using a temperature of 10,000K suggest that the 0.7890 \mum\  line will be slightly stronger than the 1.0927 \mum\ line and that the 1.0092 line should be detectable with more than 40\% of the strength of the 1.0927 \mum\ line.  These line strengths are not consistent with the spectral features.  We conclude that the region probed by NIR spectra before maximum light is closer to 5,000K than to 10,000K.

\subsection{Calculating Relative Abundance Ratios for Carbon and Oxygen}
The abundance ratio is the quotient of the line depth ratio divided by the line strength ratio.  O to C abundance ratios are determined separately for each of he five strongest \ci\ lines covered by our observations in each LRS spectrum. (\ci\ at 1.4543 \mum\ is not included because it falls in a wavelength region with poor atmospheric transmission).  Line depth ratios are calculated at all velocities in the range of the \oi\ line-forming region ($9-18 \times 10^3$ \kms).  The flux array from \oi\ at 0.7773 \mum\ is divided by the flux array for each of the \ci\ lines.  Wavelength intervals are not constant throughout the spectra, so we interpolate between \ci\ velocity data points to match the velocity spacing of the \oi\ data.  The line strength ratio for each pair of lines is the quotient of the estimated line strength for the \oi\ line divided by the estimated line strength for the \ci\ line (Tabel \ref{colines}, $T=$ 5,000K).  Thus we determine the O to C abundance ratios over a range of velocities corresponding to a large radial depth. 

We illustrate this procedure by calculating the O to C ratio for the three LRS spectra using the strongest lines from each ion at a specific velocity.  We measure the maximum line depths of the \oi\ lines and determine the velocities at maximum depth from the data displayed in Figure \ref{oi}. The depths of the \ci\ 1.0693 \mum\ lines are measured in each spectrum at the same velocities as for \oi.  The \ci\ line depths are found to be less than the one-sigma noise at those locations, so we substitute the noise values for the \ci\ line depths.  This substitution makes the O/C line depth ratios lower limits.  Table \ref{ar} gives the measured velocities, line depths, line depth ratios, estimated line strengths at 5,000K, the line strength ratios, and lower limits for the O/C ratios.  Lower limits for the oxygen to carbon abundance ratios at the velocities of \oi\ maximum depths are found to be 4,994 for SN 2004bw, 957 for SN 2000dn and 2,304 for SN 2002cr.

\begin{figure}
\plotone{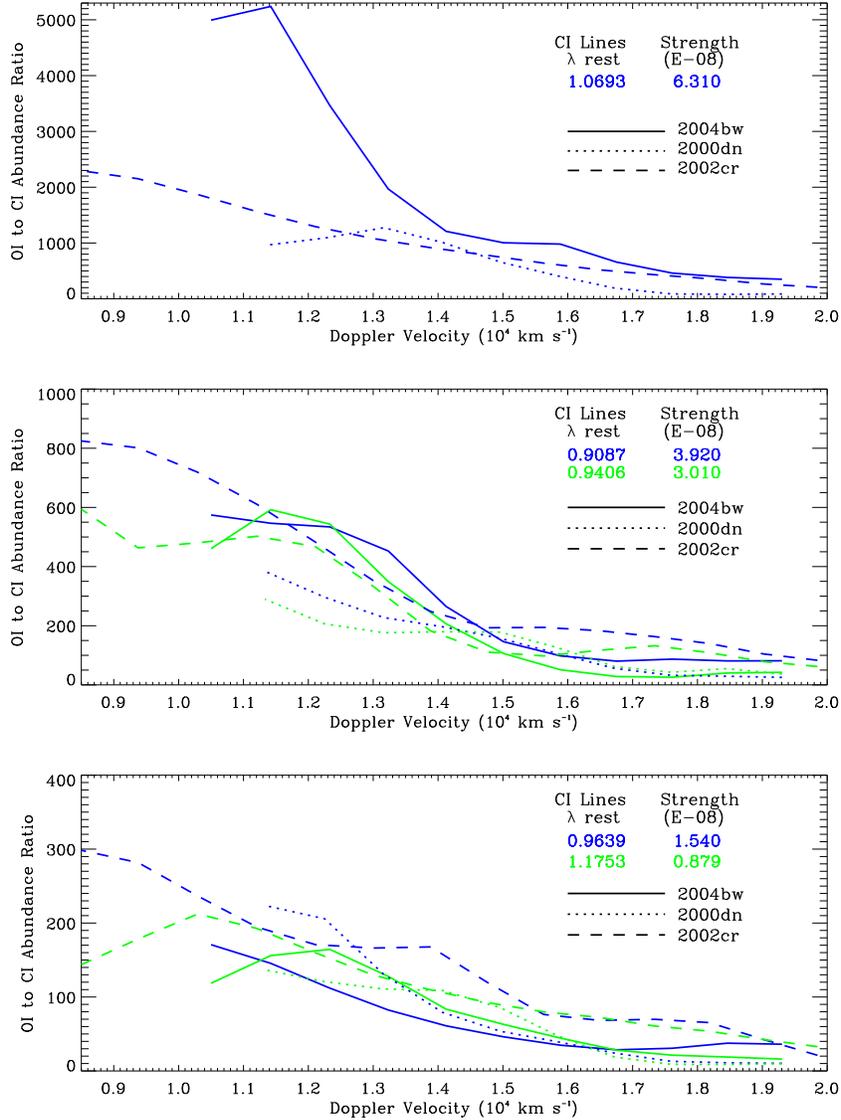}
\caption{Lower limits for estimated Oxygen to Carbon abundance ratios as a function of Doppler velocity. The \oi\ line at 0.7773 \mum\ is compared to upper limits on the five strongest \ci\ lines in the NIR. The top panel displays ratios for \oi\ to \ci\ at 1.0693 \mum, the center panel at 0.9087 and 0.9406 \mum\ and the lower panel at 0.9639 and 1.1756 \mum. Rest wavelengths of the \ci\ lines and their estimated line strengths are given in the top right corner of each panel.  We find that oxygen abundance is greater than carbon abundance by factors of $10^2-10^3$ at $\approx$ 11,000 \kms.  The lower limit for the O to C ratio remains well above unity to velocities in excess of 18,000 \kms.\label{o2c}} 
\end{figure}

Figure \ref{o2c} shows estimated oxygen to carbon abundance ratios as a function of Doppler velocities. O to C ratios have values of $10^2 - 10^3$ near the maximum measured depths for \oi\ and approach unity near 18,000 \kms\ where both O and C abundances become comparable to the one-sigma noise.  In many cases, such as the \ci\ line at 1.0693 \mum\ in the spectrum of SN 2004bw, the O/C ratio substantially exceeds unity to velocities beyond 20,000 \kms.  At nearly all velocities, the indicated abundance ratios are lower limits because no signal is detected from carbon greater than the one-sigma noise level.    

The top panel in Figure \ref{o2c} displays ratios for \oi\ to \ci\ at 1.0693 \mum, the center panel gives ratios for \ci\ at 0.9087 and 0.9406 \mum, and the lower panel presents ratios for \ci\ at 0.9639 and 1.1756 \mum. Rest wavelengths of the \ci\ lines and their estimated line strengths are given in the top right corner of each panel.  The slopes plotted in the figure do not always follow a monotonically descending path in direct proportion to the \oi\ line depth.  Instead, the shape of each plot responds to variations in local noise levels and to locations where the spectrum in the velocity region of a \ci\ line dips below the one-sigma noise.

\section{Discussion}
The amount of material remaining unburned after the explosion is an important discriminant between the leading models for SNe Ia.  Deflagration models predict that large quantities of unburned material will remain in the outer layers of the SN Ia after the explosion while Deflagration to Detonation Transition (DD) models predict that essentially all progenitor matter will experience nuclear burning.  There is agreement that within the context of standard models for stellar evolution, the composition of unburned material will be carbon and oxygen in approximately equal abundance.  Oxygen can be present in SNe Ia from either primordial material or as a product of carbon burning.  In unburned matter, C and O should both be detectable.  In C burning regions, C will have been consumed but Mg will be present as well as O.  The similarity of ionization potentials for carbon and oxygen, the small mean free path for UV photons, and features that cover several thousand \kms\ imply that carbon and oxygen will be in the same ionization state. 

We have conducted a search for evidence of carbon and oxygen using three NIR spectra from normal SNe Ia.  A strong detection is made from \oi\ at 0.7773 \mum\ with a projected velocity range of $\approx 9-18 \times 10^3$ \kms.  The line-forming region comprises about one third of the matter in SNe Ia and \oi\ extends to the outer layers.  We do not detect carbon at any velocity, in agreement with \citet{Barbon90, pah95, fisher98, Stehle05}.  If any matter from these SNe Ia contains C and O in nearly equal abundance, it must have an expansion velocity in excess of 18,000 \kms.  The upper limit on the mass of carbon beyond this velocity is: $3.6 \times 10^{-2} M_{\odot}$.  We use a DD velocity structure as a limiting case for these estimates because alternative models, such as deflagration models, produce even less mass above given expansion velocities.  

Strong, broad lines from \mg\ are also found in all spectra in our sample.  We show that the \mg\ line-forming region covers a similar velocity range to \oi, and hence a similar spatial region.  This result demonstrates that the domain of explosive carbon burning reaches the extreme outer layers of matter probed by these spectra.  

We compare estimated line strengths and measured line depths for \oi\ and \ci\ to derive lower limits on the O to C abundance ratios for the SNe Ia in our sample.  We find that the oxygen abundance is greater than the carbon abundance by factors of $10^2-10^3$ at $\approx$ 11,000 \kms, and remains well above unity to velocities in excess of 18,000 \kms.  This large imbalance and carbon depletion contradict pure deflagration models for standard progenitors that predict significant amounts of unburned C and O will remain after the explosion..  

Patchiness of the carbon and oxygen regions is sometimes invoked as a possible explanation for the lack of detection of carbon in spectra from SNe Ia, but the outer layers of published deflagration models are not patchy.  The strong \oi\ absorption features in these spectra require a large covering factor and cannot be produced by thin clumps of processed material.  

Our generalized treatment of the problem has limitations but it adds stability to our analysis and eliminates model dependencies.  The use of LTE departure coefficients introduces uncertainty in the estimated line strengths by factors of $2-3$.  The combined uncertainties in the measurement of expansion velocities are about $\pm 750$ \kms.  The estimated location of the local continuum affects measurements of line depths at less than two sigma noise levels, but is inconsequential at greater depths.   The sum of these uncertainties is not nearly large enough to change our results.

The presence of \mg\ in the same region as \oi, combined with a lack of detection of \ci\ indicate that nuclear burning has taken place in at least the outer one third of the mass in these SNe Ia and extends to more than 18,000 \kms.  We conclude that for these SNe Ia, carbon is absent at radial depths below a minimum velocity of 18,000 \kms\ because it has been explosively burnt to O and Mg.  This conclusion is in agreement with DD models for SNe Ia and contradicts pure deflagration models but it does not address the existence of C and O in SNe Ia at velocities below 9,000 \kms or above 18,000 \kms.

\citet{Quimby05} address the extreme outer layers by showing that \ion{Si}{2} is present in optical spectra of SNe Ia to velocities of $\approx$ 24,000 \kms.  Silicon in this region can only be produced from a C/O WD progenitor by explosive oxygen burning at temperatures greater than those required to burn carbon into oxygen and magnesium.  Non-standard stellar evolution processes may be able to enrich oxygen in the outer layers and reduce the amount of carbon.  This possible explanation for the lack of observed carbon is severely constrained by the presence of high temperature burning products in the same region.  Carbon abundance in the inner regions of the ejecta are constrained by \citet{Kozma05} using nebular spectra to provide very low limits on the mass of unburned material remaining in the center of SNe Ia after the explosion.  

The results of \citeauthor{Quimby05} and \citeauthor{Kozma05} combine nicely with the evidence we provide in this paper to demonstrate that nuclear burning must reach every part of the progenitor.  The lack of carbon detection in all regions of normal SNe Ia contradicts predictions of published 3-D Deflagration models that unburned C will be present at all velocities.  On the other hand, these observations are perfectly consistent with predictions of Deflagration to Detonation Transition models for SNe Ia.

\noindent{\bf Acknowledgments:}

We thank W. D. Li and M. Papenkova for providing light curve data for many of the SNe in our sample.  We thank the individuals at the IRTF for guidance and help with the observations.  In particular, Alan Tokanaga, John Rayner, Mike Cushing,  Bill Golisch,  Dave Griep, and Paul Sears have been most helpful.  We would also like to thank the TAC of the IRTF for support and instructive comments.  GHM would like to thank Mike Cushing for helpful comments.  This research is supported in part by NSF grant 0098644 and by NASA grant NAG5-7937.  CLG is supported through UK PPARC grant PPA/G/S/2003/00040.

\begin{deluxetable}{lll}
\tablecolumns{3} \tablewidth{0pc}
\tablecaption{Ionization Potential for SN Ia Explosion Products\label{ip}}
\tablehead{\colhead{Atom} & \colhead{First (eV)} & \colhead{Second (eV)}}

\startdata
C & 11.26030 & 24.38332\\
O & 13.61806 & 35.11730\\
Ne & 21.56454 & 40.96328\\
Mg & 7.64624 & 15.03528\\
Si & 8.15169 & 16.34585\\
S & 10.36001 & 23.3379\\
Ar & 15.75962 & 27.62967\\
Ca & 6.11316 & 11.87172\\
Ti & 6.8282 & 13.5755\\
Mn & 7.43402 & 15.63999\\
Fe & 7.9024 & 16.1878\\
Co & 7.8810 & 17.083\\
Ni & 7.6398 & 18.16884\\
\enddata

\end{deluxetable}

\begin{deluxetable}{llccrccc}
\tablecolumns{8} \tablewidth{7.0in}
\tablecaption{Strong Lines for Carbon and Oxygen in the Region $0.6 - 2.4$ \mum \label{colines}}
\tablehead{\colhead{Ion} & \colhead{$\lambda_{rest}$(air)} & \colhead{No. lines} &
\colhead{$\log(gf)$} & 
\colhead{Excitation} & \colhead{Line Strength} & \colhead{Line Strength}& 
\colhead{Line Strength}\\ 
 & \colhead{\mum} &\colhead{in blend \tablenotemark{a}} & &\colhead{Min. (eV)} & 
\colhead{5,000K \tablenotemark{b}} & 
\colhead{normalized \tablenotemark{c} }  & \colhead{10,000k \tablenotemark{b} }}

\startdata
   \ci &  1.0693 &   7 &   0.348 &    7.48830 &  6.307e-08 &      1.000 &  3.749e-04\\
   \ci &  0.9087 &   6 &   0.142 &    7.48830 &  3.925e-08 &      0.622 &  2.333e-04\\
   \ci &  0.9406 &   1 &   0.225 &    7.68529 &  3.008e-08 &      0.477 &  2.247e-04\\
   \ci &  0.9639 &   3 &  -0.264 &    7.48830 &  1.541e-08 &      0.244 &  9.160e-05\\
   \ci &  1.4543 &   1 &  -0.110 &    7.68529 &  1.391e-08 &      0.221 &  1.039e-04\\
   \ci &  1.1756 &   7 &   0.661 &    8.64774 &  8.792e-09 &      0.139 &  2.007e-04\\
   \ci &  0.8333 &   2 &  -0.420 &    7.68529 &  6.811e-09 &      0.108 &  5.089e-05\\
\hline
   \oi &  0.7773 &   3 &   0.324 &    9.14671 &  1.271e-09 &      1.000 &  5.177e-05\\
   \oi &  0.8446 &   3 &   0.170 &    9.52201 &  3.731e-10 &      0.294 &  2.349e-05\\
   \oi &  0.9264 &   9 &   0.690 &   10.74166 &  7.285e-11 &      0.057 &  1.889e-05\\
   \oi &  1.1290 &   9 &   0.500 &   10.98960 &  2.645e-11 &      0.021 &  9.146e-06\\
   \oi &  0.7990 &   6 &   0.280 &   10.98960 &  1.594e-11 &      0.013 &  5.511e-06\\
\hline
  \ion{C}{2} &  0.6580 &   2 &   0.118 &   14.44980 &  3.569e-15 &     88,393 &  6.844e-08\\
  \ion{C}{2} &  0.7235 &   3 &   0.330 &   16.33422 &  7.330e-17 &      1,815 &  1.252e-08\\
  \ion{C}{2} &  1.8905 &   2 &   0.258 &   19.49586 &  4.038e-20 &      1.000 &  2.705e-10\\
  \ion{C}{2} &  1.7846 &   3 &   0.550 &   20.15184 &  1.726e-20 &      0.427 &  2.474e-10\\
  \ion{C}{2}  &  0.9903 &   3 &   1.010 &   20.95206 &  7.768e-21 &      0.192 &  2.819e-10\\
\hline
  \ion{O}{2} &  0.6695 &   2 &  -0.594 &   23.44307 &  5.962e-25 &      2,155 &  3.897e-13\\
  \ion{O}{2} &  2.1085 &   2 &  -1.690 &   25.66290 &  2.766e-28 &      1.000 &  2.376e-15\\
  \ion{O}{2} &  1.3811 &   3 &  -1.730 &   25.66290 &  2.522e-28 &      0.912 &  2.167e-15\\
  \ion{O}{2} &  0.6898 &  12 &   0.230 &   28.70803 &  1.960e-29 &      0.071 &  5.770e-15\\
\enddata

\tablecomments{1995 Atomic Line Data (R.L. Kurucz and B. Bell) Kurucz CD-ROM No. 23. Cambridge, Mass.: Smithsonian Astrophysical Observatory}

\tablenotetext{a}{Where the number of lines in the blend exceeds 1, the indicated wavelength is 
a mean value, weighted by oscillator strength, for two or more lines.}

\tablenotetext{b}{The estimated line strength is computed using $(gf)*10^{-\Theta (eV)}$ where $
\Theta = 5040/T$.}  

\tablenotetext{c}{The normalized line strength is the 5,000K value divided by the strongest line for each species.}

\end{deluxetable}

\begin{deluxetable}{llccrccc}
\tablecolumns{8} \tablewidth{7.0in}
\tablecaption{Strong Lines ofr \mg\ and \ca\ in the Region $0.6 - 2.4$ \mum \label{mgcalines}}
\tablehead{\colhead{Ion} & \colhead{$\lambda_{rest}$(air)} & \colhead{No. lines} &
\colhead{$\log(gf)$} & 
\colhead{Excitation} & \colhead{Line Strength} & \colhead{Line Strength}& 
\colhead{Line Strength}\\ 
 & \colhead{\mum} &\colhead{in blend \tablenotemark{a}} & &\colhead{Min. (eV)} & 
\colhead{5,000K \tablenotemark{b}} & 
\colhead{normalized \tablenotemark{c} }  & \colhead{10,000k \tablenotemark{b} }}

\startdata
 \mg &  0.9227 &   2 &   0.270 &    8.65529 &  3.511e-09 &      1.000 &  8.086e-05\\
 \mg &  1.0927 &   3 &   0.020 &    8.86425 &  1.216e-09 &      0.346 &  3.568e-05\\
 \mg &  0.7890 &   3 &   0.650 &   10.00000 &  3.715e-10 &      0.106 &  4.074e-05\\
 \mg &  0.8228 &   2 &   0.030 &   10.00000 &  8.912e-11 &      0.025 &  9.772e-06\\
 \mg &  1.0092 &   3 &   1.020 &   11.63047 &  1.979e-11 &      0.006 &  1.440e-05\\
 \hline
 \ca &  0.8579 &   3 &  -0.362 &    1.70005 &  8.401e-03 &     13,944 &  6.042e-02\\
 \ca &  1.1876 &   2 &   0.300 &    6.46831 &  6.025e-07 &      1.000 &  1.096e-03\\
 \ca &  0.8921 &   3 &   0.729 &    7.05003 &  4.193e-07 &      0.696 &  1.499e-03\\
 \ca &  0.8235 &   3 &   0.621 &    7.51535 &  1.111e-07 &      0.184 &  6.812e-04\\
 \ca &  0.9906 &   2 &   0.072 &    7.51535 &  3.137e-08 &      0.052 &  1.924e-04\\
\enddata

\tablecomments{1995 Atomic Line Data (R.L. Kurucz and B. Bell) Kurucz CD-ROM No. 23. Cambridge, Mass.: Smithsonian Astrophysical Observatory}

\tablenotetext{a}{Where the number of lines in the blend exceeds 1, the indicated wavelength is 
a mean value, weighted by oscillator strength, for two or more lines.}

\tablenotetext{b}{The estimated line strength is computed using $(gf)*10^{-\Theta (eV)}$ where $
\Theta = 5040/T$.}  

\tablenotetext{c}{The normalized line strength is the 5,000K value divided by one of the strongest lines for each species.}

\end{deluxetable}

\begin{deluxetable}{llll}
\tablecolumns{4} \tablewidth{0pc}
\tablecaption{Lower Limits on Abundance Ratios for \oi\ at 0.7773 \mum\ to \ci\ at 1.0693 \mum\label{ar}}
\tablehead{\colhead{} & \colhead{SN 2004bw} & \colhead{SN 2000dn} & \colhead{SN 2002cr}}

\startdata

Velocity at \oi\ max. depth & 10,501 & 11,331 & 8,447\\
\oi\ max depth & $0.711 \pm 0.05$ & $0.543 \pm 0.08$ & $0.747 \pm 0.06$\\
\ci\ depth at same vel.\tablenotemark{a} & 0.007 & 0.028 & 0.016\\
Line depth ratio ($\frac{O}{C}$)& 100.5 & 19.3 & 46.4\\
\ci\ line strength estimate & $6.307 \times 10^{-8}$ & $6.307 \times 10^{-8}$ & $6.307 \times 10^{-8}$\\
\oi\ line strength estimate  & $1.271 \times 10^{-9}$ & $1.271 \times 10^{-9}$ & $1.271 \times 10^{-9}$\\
Line strength ratio ($\frac{O}{C}$)& 0.02015 & 0.02015 & 0.02015\\
O to C abundance ratio \tablenotemark{b} & 4,994 & 957 & 2,304\\

\enddata
\tablenotetext{a}{Measured \ci\ line depth at this velocity does not exceed one-sigma noise, so noise values are substituted for line depths.}
\tablenotetext{b}{Abundance Ratio = Line Depth Ratio/Line Strength Ratio.}
\end{deluxetable}

\end{document}